# Dominant spin Hall torque and negligible orbital Hall torque in α-W/ferromagnet heterostructures with artifacts-free angular momentum detectors

Taiyang Zhang,[1,2] Lujun Zhu,[3] Qianbiao Liu,[1] Lijun Zhu[1,2]*

[1] *State Key Laboratory of Semiconductor Physics and Chip Technologies, Institute of Semiconductors, Chinese Academy of Sciences, Beijing 100083, China*

[2] *Center of Materials Science and Optoelectronics Engineering, University of Chinese Academy of Sciences, Beijing 100049, China*

[3] *College of Physics and Information Technology, Shaanxi Normal University, Xi'an 710062, China*

**ljzhu@semi.ac.cn*

α-phased W was theoretically predicted to have a negative spin Hall conductivity and a positive orbital Hall conductivity at the same time, leaving the physical origin of the current-induced torque a critical open question. Here, we develop two angular momentum detectors of Cu/Ni/Cu and Cu/$Fe_{60}Co_{20}B_{20}$/Cu that are free of artifacts torques (e.g., self-induced torque and spin-vorticity torque) and clarify that the spin-orbit torque contributed by *α*-W remains negative and predominantly from the spin Hall effect in the entire thickness regime. The damping-like torque exhibits a monotonic decay as the W thickness increases above 5 nm likely due to the thickness-dependent structural evolution and resistivity reduction of the W layer. The negative torque in the entire thickness regime suggests negligible orbital Hall torque from *α*-W. This result is consistent with the theory that the orbital Hall effect from simplified band structure calculations is not a non-local angular momentum source.

Spin-orbit torques (SOTs) have attracted considerable interest as a pivotal mechanism for fast and energy-efficient electrical manipulation of magnetization in non-volatile magnetic memory and programmable computing technologies [1–6]. Since the discovery of the SOTs, the generation of SOTs has been predominantly attributed to spin currents arising directly from the spin Hall effect (SHE) [1,6–10] or the interfacial spin-orbit coupling (SOC) effects [2,10–14]. Recently, whether the orbital Hall conductivity from simplified band structure calculations contributed to SOTs has become a heated debate [15–20]. Many works have suggested that orbital polarization is strongly localized and cannot participate in non-local transport effects [21–30], presenting strong challenge to the concepts of orbital current and orbital Hall torque widely used in the past five years.

W is of particular technological importance for efficiently generating torques and switching magnetic random access memories (SOT-MRAMs)[31-33]. However, it has remained an open question as to the mechanism of the SOTs in W devices. As shown in Fig. 1(a), first-principles calculations predicted a spin Hall conductivity ($\sigma_{SH}$) of $-1.6\times10^5$ $(\hbar/2e)$ $\Omega^{-1}$ $m^{-1}$ and a 6 times greater orbital Hall conductivity of $9.3\times10^5$ $(\hbar/2e)$ $\Omega^{-1}m^{-1}$ in α-W [34]. Recently, a spin-torque ferromagnetic resonance (ST-FMR) experiment estimated a large positive torque in α-W/Ni bilayers when W is thicker than 5 nm by ignoring any contributions from the Ni [17]. However, we recently found that a Ni layer may generate significant ST-FMR signals due to the SHE and *inversion symmetry breaking* of Ni [25,35], making the single-layer Ni a contaminated detector for externally injected angular momentum. Given the critical importance of W for the SOT-MRAMs [31-33], it has been urgently required to clarify the mechanism of the torque in W-based devices.

In this letter, we fabricated artifacts-free angular momentum detectors of Cu/Ni/Cu and Cu/$Fe_{60}Co_{20}B_{20}$/Cu with *preserved-inversion-symmetry* and negligible self-generated torque and spin-vorticity torque. With the artifacts-free detectors, we clarify that the spin-orbit torque contributed by W is negative and dominated by the SHE in the W thickness range of beyond 5 nm. We also observe a monotonic decay in the dampinglike torque at large W thicknesses, which is due to the reduction of resistivity of the *α*-W and irrelevant to any positive orbital Hall effect.

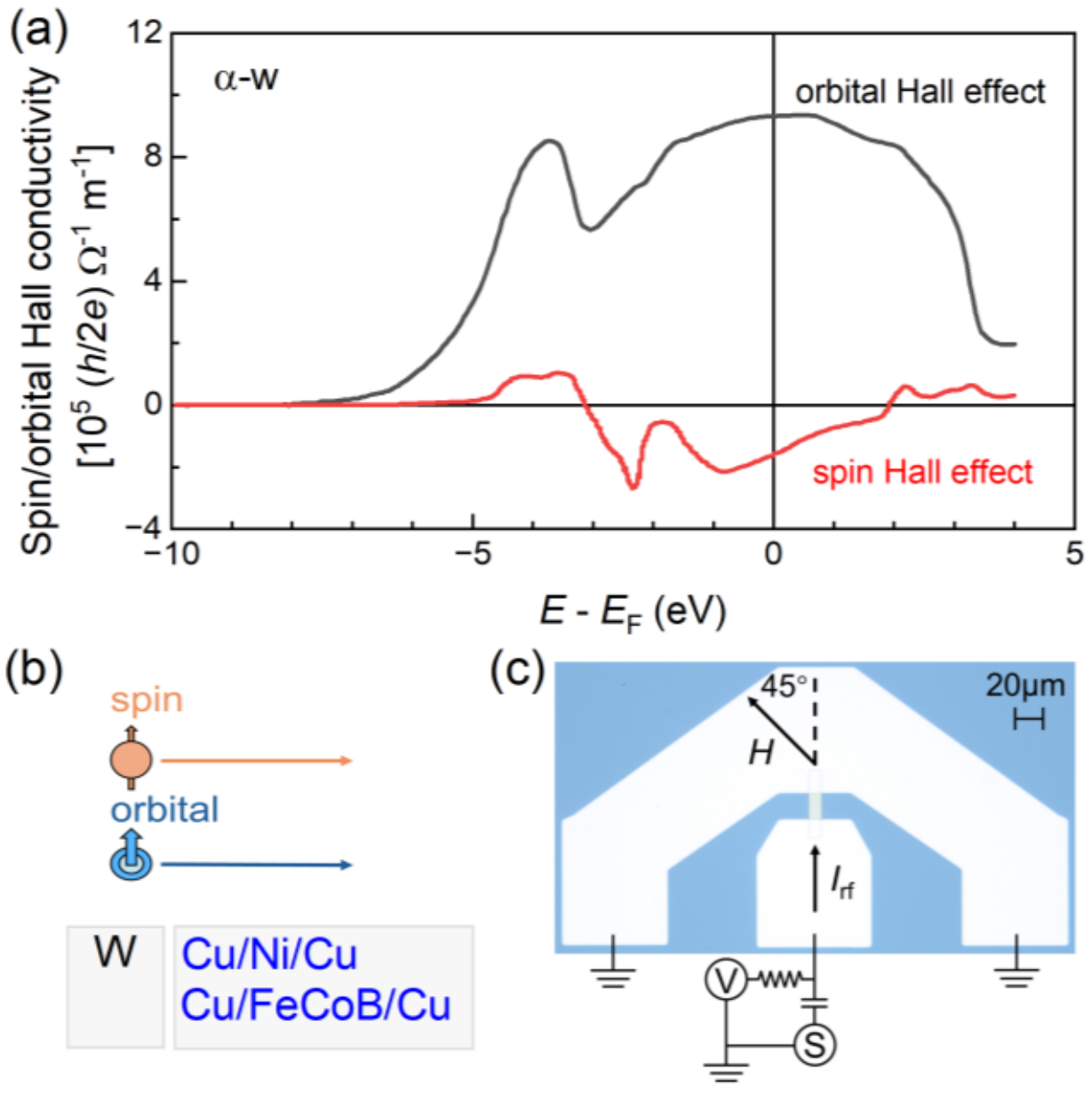

**Figure 1.** (a) Spin and orbital Hall conductivities from first-principles band structure calculations of α-W (replotted from [34]). (b) Potential spin and orbital transport from W into the Cu/Ni/Cu and Cu/ $Fe_{60}Co_{20}B_{20}$/Cu. (c) Optical microscopy image and measurement geometry of a ST-FMR device.

***Structural characterizations.*** We first sputter-deposited in-plane magnetized multilayers of Cu 2.4/Ni 4.8-8/Cu 2.4/W 0-20 (Fig. 1(b)) at room temperature under argon pressure of 2 mTorr on thermally oxidized silicon substrates. Here, the numbers are layer thicknesses in nanometers. Ni is chosen as the torque detector since it is widely claimed to be an outstanding orbital-spin converter [15–20]. The 2.4 nm Cu spacers below and above the Ni are used to maintain the inversion-symmetry and diminish the self-induced spin Hall torque and spin-vorticity torque within the Ni (see below). Each multilayer is protected by a MgO 2/Ta 1.5 nm bilayer that is fully oxidized upon exposure to the atmosphere [36]. The sputtering system has the ultrahigh base vacuum of <1 × $10^{-9}$ Torr. To avoid inter-diffusion and magnetic dead layer, each layer was deposited with a very low rate, i.e., 0.021 nm/s for Cu, 0.014 nm/s for W, 0.024 nm/s for Ni, 0.002 nm/s for MgO, 0.009 nm for Ta. We grow the Cu/Ni/Cu detectors always on Si/$SiO_2$ substrates to best prevent any variation of the Ni due to the varying thicknesses of the W layers. This recipe is advantageous since in the conventional W/Ni structure the Ni was directly grown on the surfaces of W layers with varying thicknesses. These multilayers are patterned into 10 × 20 $\mu m^2$ ST-FMR devices (Fig. 1(c)) by photolithography and ion milling, followed by the deposition of Ti 5/Pt 150 contacts.

In general, the sputter-deposited W films can form amorphous phase, $\alpha$ phase (bcc, thermodynamically stable), and $\beta$ phase (A15, metastable), depending on the layer thickness, base vacuum, and annealing condition [37-39]. The structures of the W layers in this Letter are characterized by the energy-dispersive x-ray spectroscopy (EDS) mapping, x-ray diffraction, and the high-resolution high-angle annular dark-field scanning transmission electron microscopy (HAADF-STEM) imaging. We verify the layer thicknesses and reasonably sharp interfaces of our devices by the EDS mapping in Fig. 2(a). We further performed x-ray $\theta$-$2\theta$ diffraction measurements with the x-ray beam diameter of 5 mm to characterize the structures of the W layers with different thicknesses. As plotted in Fig. 3(b), all the W layers exhibit only the $\alpha$ (110) peak but no indication of the $\beta$ phase. In the $2\theta$ range of 35°-45°, the $\beta$ phase, if present, should have two almost equally strong (211) at 43.9° and (210) peak at 39.9° along with a relatively low (200) peak at 35.5°. This result reveals that the W layers is dominated by the $\alpha$ phase. From the HAADF-STEM imaging in Fig. 2(c), the 20 nm W layer has bright bcc pillar-like grains with the lattice plane spacing of 0.22 nm for the (110) planes and 0.32 nm for the (100) planes, which are consistent with what would be expected for $\alpha$-W. From the HAADF-STEM image of the 5 nm W device in Fig. 2(d) and Fig. S1 in the Supplementary Materials [40], we reaffirm that the 5 nm W is also dominated by the $\alpha$ phased, which is not surprisingly given the previous reports of 5 nm α-W in as-deposited films [37]. The W in the previous W/Ni experiment [17] was also introduced as α-phased in the thickness regime of 5-20 nm.

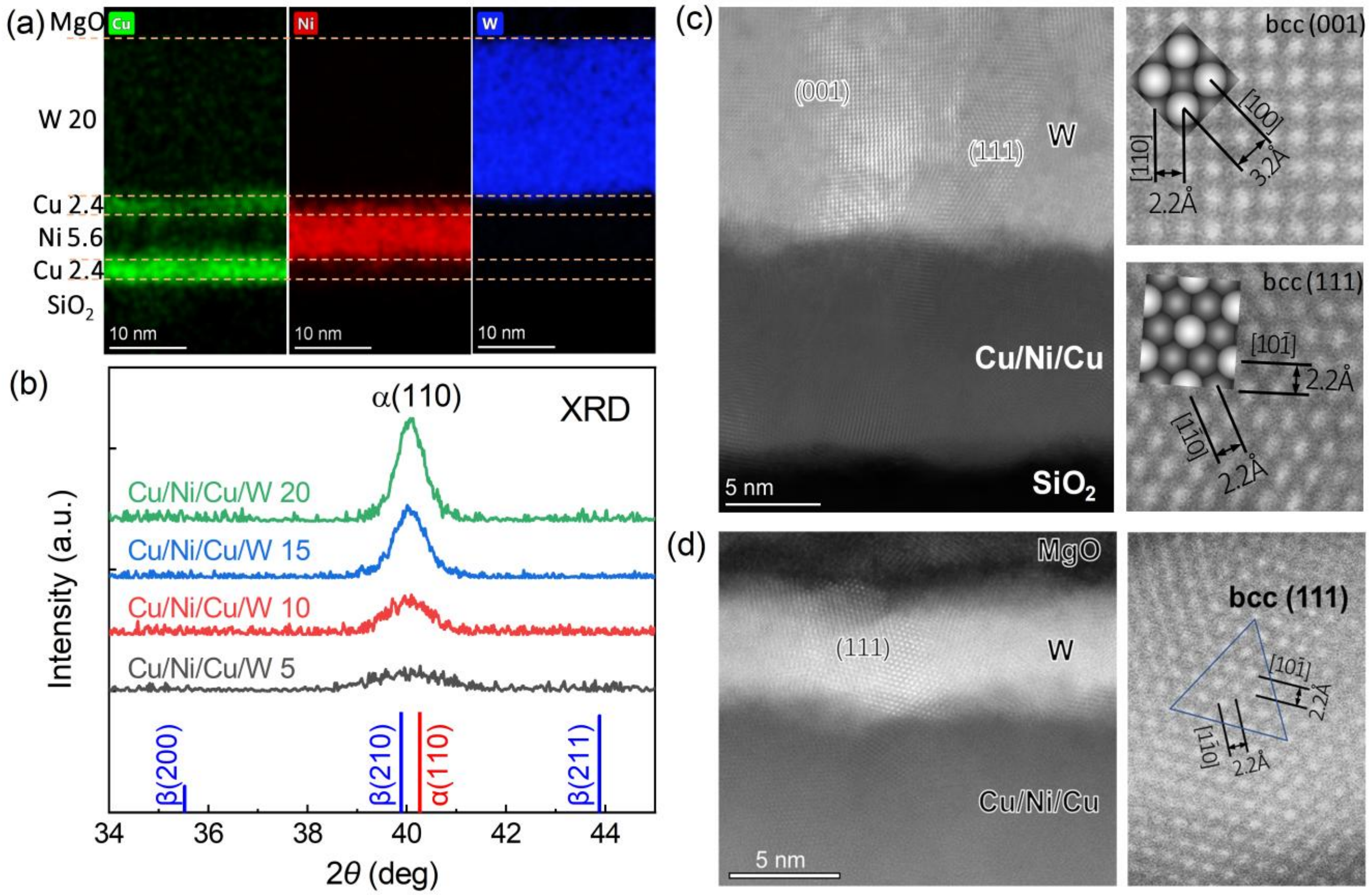

Fig. 2. (a) Cross-sectional EDS elemental mapping of the Cu 2.4/Ni 5.6/Cu 2.4/W 20 device. (b) X-ray diffraction pattern of the Cu 2.4/Ni 4.7/Cu 2.4/W $d$ with different W thicknesses $d$, suggesting all the W layers are $\alpha$ phased. The red and blue lines indicate the $2\theta$ angles and relative intensities of the diffraction peaks of $\alpha$-W and $\beta$-W from the joint committee on powder diffraction system (JCPDS) file data. (c) HAADF-STEM image of the Cu 2.4/Ni 5.6/Cu 2.4/W 20 sample for ST-FMR analysis. (d) HAADF-STEM image of the Cu 2.4/Ni 5.6/Cu 2.4/W 5 sample.

***Spin-torque magnetic resonance measurements.*** As shown in Fig. 1(c), the dampinglike torques of the Cu/Ni/Cu/W multilayers are determined using the standard three-terminal ST-FMR technique [7,41,42]. During the measurements, the rf current is sourced via a bias Tee into the magnetic strip symmetrically from the center electrode and flows out from two ground electrodes. The longitudinal rectified voltage ($V_{mix}$) due to the interplay of the magnetoresistance and SOTs is recorded from the bias Tee using a voltmeter as a function of the swept in-plane field ($H$) along the azimuth angle ($\varphi$) of 45° relative to the rf current direction. As shown in Fig. 3(a), the ST-FMR spectra include symmetric ($S$) and antisymmetric ($A$) Lorentzian components [41] as given by

$$V_{\mathrm{mix}} = S\frac{\Delta H^2}{\Delta H^2 + (H - H_{\mathrm{r}})^2} + A\frac{\Delta H (H - H_{\mathrm{r}})}{\Delta H^2 + (H - H_{\mathrm{r}})^2}, \quad (1)$$

where $\Delta H$ is the half width at the half maximum of FMR and $H_{\mathrm{r}}$ the resonance field. When the $S$ and $A$ signals are contributed by the W as is the case of this work, one can define an FMR efficiency [42]

$$\xi_{\mathrm{FMR}} = \frac{S}{A}\frac{4\pi M_{\mathrm{s}} t_{\mathrm{FM}} d_{\mathrm{HM}} e}{\hbar}\sqrt{1 + 4\pi M_{\mathrm{eff}}/H_{\mathrm{r}}}, \quad (2)$$

where $e$ is the elementary charge, $\mu_0$ the permeability of the vacuum, $\hbar$ the reduced Planck's constant, $M_{\mathrm{s}}$ the saturation magnetization, $t_{\mathrm{FM}}$ the thickness of the magnetic layer, $d_{\mathrm{HM}}$ the thickness of the heavy metal layer that generates angular momentum (the W layer), and $4\pi M_{\mathrm{eff}}$ the effective magnetization. $4\pi M_{\mathrm{eff}}$ is extracted from the dependence of $H_{\mathrm{r}}$ on the rf frequency ($f$) following Kittel's equation [43]

$$f = \frac{\gamma}{2\pi}\sqrt{H_{\mathrm{r}}(H_{\mathrm{r}} + 4\pi M_{\mathrm{eff}})}, \quad (3)$$

where $\gamma$ is the gyromagnetic ratio. As plotted in Fig. 2(b), $M_{\mathrm{s}}$ is then estimated to be 232-311 emu/cm$^3$ for the different devices from the linear fit of $4\pi M_{\mathrm{eff}}$ vs $1/t_{\mathrm{Ni}}$ [44]

$$4\pi M_{\mathrm{eff}} \approx 4\pi M_{\mathrm{s}} - 2K_{\mathrm{s}}/M_{\mathrm{s}} t_{\mathrm{FM}}, \quad (4)$$

where $K_{\mathrm{s}}$ is the interfacial perpendicular anisotropy energy density. $K_{\mathrm{s}}$ of the Cu/Ni/Cu/W devices is very low (0.07-0.19 erg/cm$^2$, Fig. S2 [40]) compared to typical values of heavy metal/ferromagnet bilayers (1-2 erg/cm$^2$)[28,45], suggesting little role of interfacial SOC [46].

As shown in Fig. 2(c), the dampinglike torque efficiency ($\xi_{\mathrm{DL}}^{j}$) and the fieldlike torque efficiency ($\xi_{\mathrm{FL}}^{j}$) were then estimated from the inverse intercept and the slope of the linear fit of $d/\xi_{\mathrm{FMR}}$ versus $1/t_{\mathrm{FM}}$ in the small-thickness regime following [7]

$$\frac{1}{\xi_{\mathrm{FMR}}} = \frac{1}{\xi_{\mathrm{DL}}^{j}}\left(1 + \frac{\hbar \xi_{\mathrm{FL}}^{j}}{4\pi M_{\mathrm{s}} t_{\mathrm{FM}} d_{\mathrm{HM}} e}\right). \quad (5)$$

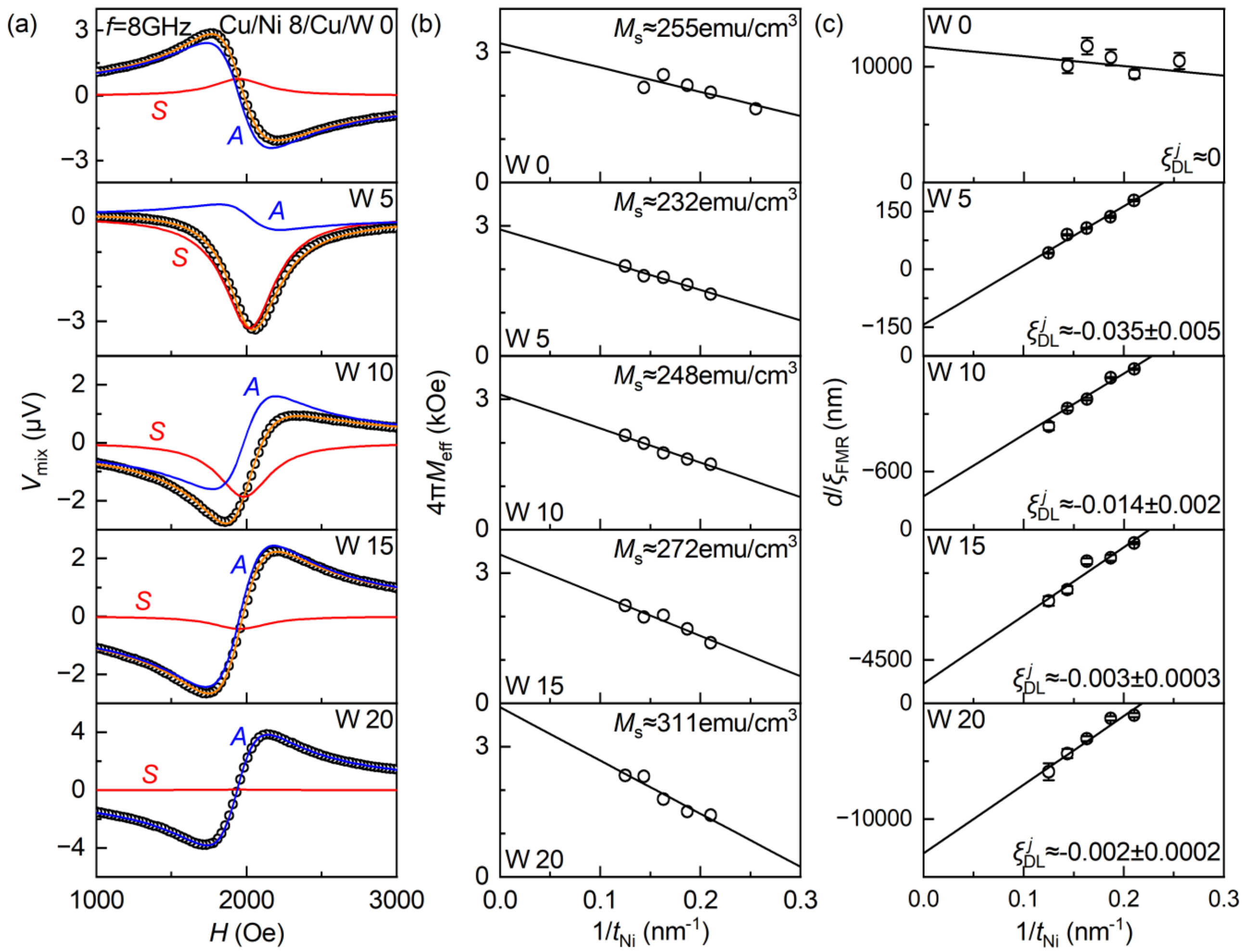

**Fig. 3.** (a) ST-FMR spectra for the Cu/Ni 8/Cu/W for the W thickness of 0 nm, 5 nm, 10 nm, 15 nm, and 20 nm (8GHz). The orange curves represent the fits of data to Eq. (1), and the red and blue curves plot the symmetric and antisymmetric components, respectively. (b) $4\pi M_{\mathrm{eff}}$ and (c) $d/\xi_{\mathrm{FMR}}$ versus the inverse Ni thickness $1/t_{\mathrm{Ni}}$ for the Cu/Ni/Cu/W samples. Some error bars that are smaller than symbols are not shown in (b) and (c). Solid lines in (b) and (c) represent the linear fits of the data to Eq. (3) and Eq. (5). The rf power is 8 dBm.

***Spin-orbit torques.*** As shown in Fig. 4(a), the Cu/Ni/Cu without a W layer has vanishingly small $\xi_{DL}^{j}$, reaffirming that the Cu under- and over-layers have successfully eliminated any self-induced torque in the Ni. $\xi_{DL}^{j}$ is -0.035 ± 0.005 for Cu/Ni/Cu/W 5 and decreases monotonically as the W thickness further increases, i.e., -0.014 ± 0.002 for Cu/Ni/Cu/W 10, -0.003 for Cu/Ni/Cu/W 15, and -0.002 for Cu/Ni/Cu/W 20. The decrease of the torque with the W thickness is essentially consistent with that of our Cu 2.4/$Fe_{60}Co_{20}B_{20}$ 2-6/Cu 2.4/W 5-15 control devices (the $Fe_{60}Co_{20}B_{20}$ was grown at a low rate of 0.013 nm/s), in which the $Fe_{60}Co_{20}B_{20}$ is widely considered sensitive only to spin current but not orbital current [47-49]. Given the widespread belief of the distinct orbit angular momentum detection effectiveness of Ni (highly sensitive) and the $Fe_{60}Co_{20}B_{20}$ (insensitive) in the field of orbitronics [47-49], the consistency of the torque efficiencies of the Cu/Ni/Cu/W and the Cu/$Fe_{60}Co_{20}B_{20}$/Cu/W suggests that orbit torque in the Cu/Ni/Cu/W is negligible and that the decrease of the negative torque at large W thicknesses is due to the decrease of the spin Hall torque rather than an increasingly strong positive orbital torque from the W.

The lack of a detectable positive orbital torque in the Cu/Ni/Cu/W cannot be attributed to the thin Cu spacer layers. The 2.4 nm Cu spacer between the Ni and the W is not expected to substantially attenuate or magnify the potential spin and orbital currents according to previous reports that Cu should have very long spin and orbital diffusion lengths (9-100 nm)[50-52] and negligible spin and orbital Hall conductivities [53]. The small, negative spin Hall torque from the 20 nm α-W is further ensured by independent harmonic Hall voltage measurement on an additional FePt 5/α-W 20 device (Fig. S3 [40]), in which the α-W is contacted directly, without a Cu spacer, by the magnetic detector FePt with strong spin-orbit coupling and negligible self-generated torque [54].

We further discuss that the interesting thickness dependence of the torque ($\xi_{DL}^{j} = \sigma_{SH}\rho_W$ [9]) is most likely due to the reduction of resistivity upon increasing the thickness. In Fig. 4(b), we plot the resistivity ($\rho_W$) of the W estimated from control Hall bar devices of $\alpha$-W layers grown on 1 nm Cu seeded Si/$SiO_2$ substrates (see Fig. S5 in the Supplementary Materials [40] for the XRD data). $\rho_W$ decreases from 280 μΩ cm at $d$ = 5 nm to 75 μΩ cm at $d$ = 20 nm likely due to a thickness-dependent evolution of disorder and grain boundary scattering. The influence of interfacial electron scattering should be weak since mean-free path of the resistive $\alpha$-W (≤1.1 nm [55]) is much less than the layer thickness. The resistivity of 75-280 μΩ cm are within the typical range of the resistivity of $\alpha$-W layers [37,38]. The negative, small $\xi_{DL}^{j}$ is also consistent with the negative, low spin Hall ratio of $\alpha$-W [31,56,57] but in contrast to pure $\beta$-W that has large spin Hall ratio of ⩾-0.3 [31,57] and resistivity of ⩾300- 800 μΩ cm for $\beta$-W [31,37,38]. Thus, we conclude that the damping-like torque of the α-W devices is dominated by the spin Hall torque. The observed monotonic decay of $\xi_{DL}^{j}$ without turning positive in the large-thickness α-W samples aligns with previous reports [31,57].

Finally, we stress that there should be little spin-vorticity torque in our Cu/Ni/Cu/W and Cu/$Fe_{60}Co_{20}B_{20}$/Cu/W devices. In the proposal of spin-vorticity torque for NM/FM bilayers [22,27], the orbital angular momentum due to the resistivity gradient is localized such that only the resistivity gradient within the FM layer may have direct relevance to the torque experienced by the FM. The spatial separation of the Ni and FeCoB layers, which have thicknesses much greater than its mean-free path (1.6 nm @25 μΩ cm for Ni [55] and ~1 nm @150 μΩ cm for FeCoB [58]), from the W from by the 2.4 nm Cu should have essentially eliminate any impact of the W on the resistivity gradient of the Ni interface, which further assures the inverse symmetry of the Cu/Ni/Cu. We have verified from finite-element modeling that the current densities in the top and bottom Cu layers remain identical (relative difference is of the order of magnitude of $10^{-5}$, see Note 2 in the supplementary Materials [40]). In addition, spin-vorticity torque should be effective only when the resistivity of the normal metal was far smaller than that of the ferromagnet [27], which is not the case for the Ni/Cu interface in which both the Ni and Cu are similarly conductive. The linear scaling of $1/\xi_{FMR}$ vs $1/t_{Ni}$ for the Cu/Ni/Cu/W in Fig. 2(c) also suggests the robustness of the torque against the variation of the thickness (resistivity non-uniformity) of the Ni. We have also measured negligible torque from our Cu 0-5/FePt (which will be published elsewhere), Al 0-4/FePt [26], and Ti/$Fe_{60}Co_{20}B_{20}$ devices [28], in which the normal metal layer is much less resistive than the ferromagnets. Many other groups have also measured negligible torques from Cu/Co, Ti/$Fe_{60}Co_{20}B_{20}$ devices [59], FePt/Cu [60]. We consider these experiments [26,28,59,60] suggest that the spin-vorticity torque is generally negligibly weak.

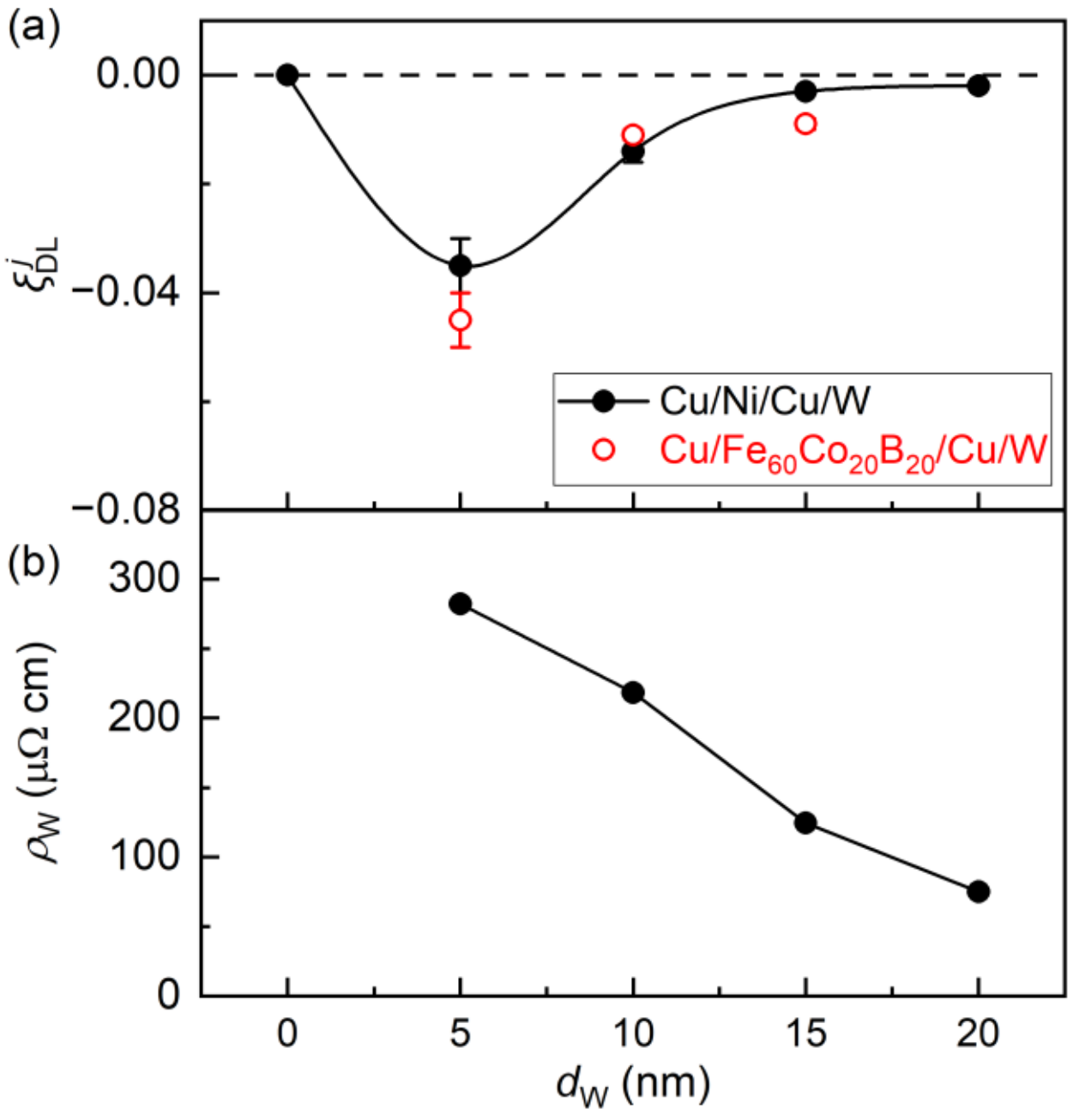

**Fig. 4.** (a) Dampinglike torque efficiency ($\xi_{DL}^{j}$) plotted as a function of the W layer thickness ($d_W$) for the Cu/Ni/Cu/W and Cu/FeCoB/Cu/W multilayers. (b) Thickness dependence of the resistivity of tungsten ($\rho_W$). Some error bars in (a) and (b) are smaller than the symbols and not shown. The lines in (a) and (b) are guides to the eyes.

***Conclusion***. Utilizing Cu/Ni/Cu and Cu/FeCoB/Cu as two angular momentum detectors free of artifacts torque (self-induced torque and spin-vorticity torque), we have demonstrated robust evidence that spin Hall torque dominates in the W-based devices and that there is negligible torque contribution from the theoretically predicted positive orbital Hall conductivity of α-W. We infer that the positive damping-like torque for α-W/Ni bilayers beyond the Ni thickness of 5 nm in previous work [17] included artifacts torque from the Ni (e.g., self-induced torque [15,25]). We also find that the damping-like torque decreases as the W thickness increases from 5 nm to 20 nm likely due to the thickness-dependent structural evolution and resistivity reduction of the W layer. The consistent absence of the orbital current torque in the W and previously reported Ta [37], Al [26], and Ti [28-30] consistently suggests negligible orbital current torque in magnetic heterostructures. This work will stimulate precise verification of orbital current and orbital Hall torque in other systems.

**Acknowledgements.** The National Key Research and Development Program of China (2022YFA1204000), the National Natural Science Foundation of China (12274405), and the Natural Science Foundation of Shaanxi Province (2024JC-YBMS-315).